\documentclass[aps,pre,amssymb,10pt]{revtex4}
\usepackage{graphicx}
\usepackage{bm}
\usepackage{color}
\usepackage{amsmath}
\usepackage[T1]{fontenc}
\usepackage[utf8]{inputenc}
\usepackage{float}

\begin{document}

\title{Nonlinearity management of matter-wave vector solitons of
Bose--Einstein condensates in two dimensions}

\author{F. Kh. Abdullaev$^{1,2}$}
\author{J. S. Yuldashev$^{1,2}$}
\author{M. \"{O}gren$^{3,4}$}
\affiliation{$^1$Physical-Technical Institute, Uzbek Academy of Sciences, 100084 Tashkent, Uzbekistan}
\affiliation{$^2$Institute of Theoretical Physics, National University of Uzbekistan, 100174, Tashkent, Uzbekistan}
\affiliation{$^3$School of Science and Technology, \"{O}rebro University, 70182 \"{O}rebro, Sweden}
\affiliation{$^4$HMU Research Center, Institute of Emerging Technologies, Heraklion, GR-71004, Greece}
\date{\today}

\begin{abstract}
The evolution of matter-wave vector solitons in  two dimensions under nonlinearity management is studied. The averaged over strong and rapid
modulations in time of the inter-species interactions vector Gross-Pitaevskii equation is derived. The
averaging gives the appearance of the effective nonlinear quantum pressure depending on the population of the
other component. Using this system of equations,
the dynamical stabilization of the two-dimensional vector matter-wave
solitons under the management of the mean-field nonlinearity is investigated.
It is shown that the nonlinear quantum pressure arrests the collapse in the averaged system, giving rise to stable vector solitons. Full numerical simulations of the original, time-dependent system of Gross-Pitaevskii equations confirm that the management strongly prolongs the lifetime of these solitons compared to the unmanaged case.
\end{abstract}

\maketitle
\section{Introduction}\label{sec_intro}

The properties of an atomic Bose-Einstein condensate (BEC) under
modulation in time of a scattering length (i.e. Feshbach resonance management) have attracted large attention the last years~\cite{PerManSol,AGKT}. The existence of two-dimensional bright
solitons, in one and two component attractive condensates~\cite{Kevrekidisbook,Malomrev,SU,Abd03,Montesinos,PG}, in media with competing nonlinearities~\cite{AOY}, long lived Bloch oscillations of gap solitons~\cite{BKS}, Faraday waves~\cite{Staliunas}, existence of stable compactons~\cite{AKS}, generation of intensity dependent gauge fields~\cite{Santos1,Santos2}, and other interesting phenomena, has been predicted and observed.
We are here interested in the study of two-component BEC. In distinction from the one-component (so-called scalar) case,
effects of Feshbach resonance management can then be on inter- and
intra-species scattering lengths, leading to different scenarios of the collapse arrest and richer varieties of stabilized localized states.
Modulational instability and collapse phenomena for the vector NLSE have been analyzed in the pioneering work by Berkhoer and Zakharov~\cite{BZ}. Note that dynamical stabilization of collapse in two-component BECs by quantum fluctuations recently have been suggested~\cite{Petrov,PA}. This scheme is applied to the intra-component repulsive and inter-component attractive BEC, with a small residual mean-field attraction in order of the strength of the quantum fluctuations. Here we will instead consider the case when all interactions are attractive and then the approach of~\cite{Petrov} is not applicable.

Mathematically the problem is reduced to the investigation of dynamically stabilized localized states in two coupled nonlinear
Schr\"odinger equations (NLSE) with nonlinearity management (NM) of the
self- and cross-phase modulation terms.

We will consider the dynamics of  nonlinear vector matter-waves under periodic modulations in time of the mean-field nonlinearity. In a two-component condensate the management can act on the inter-species scattering length alone, leaving the intra-species interactions untouched. Note that in the work~\cite{PG}
the possibility of the stabilization of two-dimensional vector NLSE has been investigated mainly by numerical simulations. The one-dimensional case of this work was studied in~\cite{AOY_PLA}.
Here both analytical and numerical analysis will be performed.
We will derive the averaged vector Gross-Pitaevskii Equation~(GPE), which is called vector NLSE in optics, with effectively nonlinear dispersion terms (corresponding to the appearance of the nonlinear quantum pressure). Based on this averaged equation, we analyse the conditions of the existence of bright solitonic states,
and we study the properties of these solutions. By the numerical simulations of the original vector GPE with nonlinearity management, we verify the analytical predictions, including the averaged equation approach, and analyze the results further than the analytical predictions.

The article is organised as follows. In Section~\ref{sec_2} we introduce the model, recall the collapse criterion, and derive the system of GPEs equations averaged over strong and rapid modulations of the inter-species interaction. Section~\ref{sec_3} applies a variational analysis to the averaged
system: we obtain the effective Lagrangian, reduce the dynamics to a
two-degree-of-freedom mechanical problem, and locate and classify the fixed points for symmetric and asymmetric configurations. Section
\ref{sec_Numerical_results} compares the variational predictions with numerical solutions of both the averaged and the original GPE systems, quantifies the accuracy of the Gaussian ansatz. Section~\ref{Sec_Conclusion} concludes the article.

\section{Model}\label{sec_2}

Two-component ($\psi_{1,2}$) untrapped BECs in two dimensions are described by
the system of coupled Gross--Pitaevskii equations
\begin{eqnarray}\label{eq:dimensional}
i\hbar\psi_{1,\tilde{t}} &=&-\frac{\hbar^2}{2m_1}\tilde{\nabla}^2 \psi_1
+ (\tilde{g}_{11} |\psi_1|^2 + \tilde{g}_{12}|\psi_2|^2)\psi_1 , \nonumber \\
i\hbar\psi_{2,\tilde{t}} &=&-\frac{\hbar^2}{2m_2}\tilde{\nabla}^2\psi_2
+ (\tilde{g}_{22} |\psi_2|^2 + \tilde{g}_{21}|\psi_1|^2)\psi_2 .
\end{eqnarray}
Here $\tilde{g}_{ij}=2\sqrt{2\pi}\hbar^{2} a_{ij}/(m l_z)$ are the effective
two-dimensional coupling constants obtained by integrating out a tightly
confined transverse direction, $a_{ij}$ are the inter- and intra-species
$s$-wave scattering lengths, and $l_z=\sqrt{\hbar/(m\omega_z)}$ is the
transverse oscillator length associated with the trap frequency $\omega_z$.
We consider equal masses $m=m_1=m_2$ and introduce the dimensionless variables
\begin{equation}\label{eq:dim_units}
(x,y)=(\tilde{x},\tilde{y})/l_z,\quad t=\tilde{t}\,\omega_z,\quad
u=l_z\psi_1,\quad v=l_z\psi_2,\qquad
g_{ij}=\frac{2\sqrt{2\pi}\,a_{ij}}{l_z} .
\end{equation}
With this choice the norms
\begin{equation}\label{eq:norms}
N_{u}=\int_{\mathbb{R}^2}|u|^2d^2x,\qquad
N_{v}=\int_{\mathbb{R}^2}|v|^2d^2x,
\end{equation}
are precisely the numbers of atoms in the two components, and
Eqs.~(\ref{eq:dimensional}) become
\begin{eqnarray}\label{eq:original_model}
i u_t+ \tfrac{1}{2}\nabla^2 u - (g_{11}|u|^2 +g_{12}|v|^2)u&=&0 , \nonumber\\
i v_t+ \tfrac{1}{2}\nabla^2 v - (g_{22}|v|^2+ g_{21}|u|^2)v&=&0 ,
\end{eqnarray}
with $g_{12}=g_{21}=\gamma$ and $\nabla^2=\partial_x^2+\partial_y^2$.
With this sign convention attractive interactions correspond to $g_{ij}<0$.
In nonlinear optics $t$ plays the role of the propagation coordinate and
$u,v$ denote two mutually incoherent beams.

The conserved Hamiltonian of the autonomous system (i.e. $\gamma=$~const) is then
\begin{equation}\label{eq:Hamiltonian}
H=\int_{\mathbb{R}^2}\Big( \tfrac{1}{2}|\nabla u|^2 + \tfrac{1}{2}|\nabla v|^2
+\tfrac{1}{2} g_{11}|u|^4+\tfrac{1}{2}g_{22}|v|^4 + \gamma|u|^2|v|^2 \Big)d^2x .
\end{equation}

\subsection{Collapse criterion}\label{sec:virial}

Two dimensions is the critical dimension for the cubic NLSE, and the vector
system~(\ref{eq:original_model}) is subject to wave collapse~\cite{Vlasov}. Introducing the
mean square radius
\begin{equation}
\bar{X}(t)=\frac{1}{N}\int_{\mathbb{R}^2}\rho^2\left(|u|^2+|v|^2\right)d^2x ,
\qquad \rho^2=x^2+y^2,\quad N=N_u+N_v ,
\end{equation}
the virial (moment) theorem~\cite{Vlasov,McKin,Bang,Veks} for
Eqs.~(\ref{eq:original_model}) gives
\begin{equation}\label{eq:virial}
\frac{d^2\bar{X}}{dt^2}=\frac{4H}{N} .
\end{equation}
Consequently $\bar{X}(t)\le \bar{X}(0)+\dot{\bar{X}}(0)t+2Ht^2/N$, so that
$H<0$ forces $\bar{X}$ to vanish in finite time, i.e. the solution blows up.
For radially symmetric Gaussian initial wave packets of width $w$
\begin{equation}\label{eq:gauss_ic}
u,v=\left(\frac{N_{i}}{\pi w^2}\right)^{1/2}e^{-\rho^2/(2w^2)} ,
\end{equation}
Eq.~(\ref{eq:Hamiltonian}) yields
\begin{equation}\label{eq:H0}
H(0)=\frac{1}{2w^2}\Big( N+\frac{1}{2\pi}\sum_{i,k=1}^{2}g_{ik}N_iN_k \Big),
\qquad (N_1,N_2)\equiv(N_u,N_v).
\end{equation}
Hence, collapse of such packets is guaranteed whenever the quadratic form
$Q=g_{11}N_u^2+g_{22}N_v^2+2\gamma N_uN_v$ satisfies $Q<-2\pi N$.

Equation~(\ref{eq:H0}) uses the same width for both components, which is
unnecessarily restrictive. Minimising $H$ over the two-parameter Gaussian
family $u,v\propto\exp(-\rho^2/2w_{1,2}^2)$ gives
\begin{equation}\label{eq:Hw1w2}
H(w_1,w_2)=\frac{a_1}{w_1^2}+\frac{a_2}{w_2^2}+\frac{a_{12}}{w_1^2+w_2^2},
\qquad
a_i=\frac{N_i}{2}+\frac{g_{ii}N_i^2}{4\pi},\qquad
a_{12}=\frac{\gamma N_u N_v}{\pi},
\end{equation}
which is negative for some $(w_1,w_2)$ since $\gamma < 0$ --- and hence guarantees collapse ---
if and only if $(\sqrt{a_1}+\sqrt{a_2})^2<-a_{12}$, i.e.
\begin{equation}\label{eq:sharpNc}
 \sqrt{2\pi N_u+g_{11}N_u^2}+\sqrt{2\pi N_v+g_{22}N_v^2}
\ <\ 2\sqrt{-\gamma N_uN_v},
\end{equation}
with the optimal width ratio being $w_2/w_1=\kappa_*=(a_2/a_1)^{1/4}$. Only the products $g_{ij}N_i$ enter, as the scaling invariances of
Sec.~\ref{sec:pdescaling} require. Condition (\ref{eq:sharpNc}) is sharper than the equal-width estimate whenever the populations are unbalanced, and it will reappear in Sec.~\ref{sec:asymmetric}
as the exact existence boundary of the asymmetric managed solitons.
In the symmetric situation $g_{11}=g_{22}=g$, $N_u=N_v=N_0$ it reduces to
\begin{equation}\label{eq:Nc}
N_0>N_c=\frac{2\pi}{|g+\gamma|}, \qquad g+\gamma<0 .
\end{equation}
The exact, non-Gaussian threshold is somewhat lower (see Sec.~\ref{sec:townes}).
When at least one intra-species interaction is repulsive, $Q$ can be made negative only if~\cite{Veks}
\begin{equation}\label{eq:Veks}
\gamma\le -\sqrt{g_{11}g_{22}} ,
\end{equation}
which is the condition used below: both species are self-repulsive
($g_{11},g_{22}>0$, Set $B$ in the table~\ref{tab:sets}) while the inter-species interaction is attractive and strong enough to drive the collapse.

\subsection{Averaged equations}\label{sec:averaging}

We now let the inter-species coefficient be rapidly and strongly modulated,
\begin{equation}\label{eq:gammat}
g_{12}(t)=g_{21}(t)=\gamma(t)=\gamma_0+\gamma_1\cos(\omega t),
\qquad \gamma_1,\ \omega\sim 1/\epsilon,\ \ \epsilon\ll 1 ,
\end{equation}
where $\omega$ is the dimensionless modulation frequency. In this regime, $\gamma_1/\omega=O(1)$, 
which corresponds to the case of \emph{strong} nonlinearity management. 
Experimentally, $\gamma(t)$ is controlled through the inter-species scattering length $a_{12}$ 
by a (possibly optically induced) Feshbach resonance~\cite{Feshbach}; in optics, 
all $g_{ij}$ are modulated along the propagation coordinate~\cite{PG}.

To average over the fast modulation we follow the method
of~\cite{Pelinovskii_2,Stepanov} and perform the near-identity transformation
\begin{equation}\label{eq:transformations}
u=\bar{u}\,e^{-i\varGamma(t)|\bar{v}|^2},\qquad
v=\bar{v}\,e^{-i\varGamma(t)|\bar{u}|^2},\qquad
\varGamma_t=\gamma_1(t),\quad
\varGamma(t)=\frac{\gamma_1}{\omega}\sin(\omega t) ,
\end{equation}
which removes the strongly oscillating cubic terms. Substituting
(\ref{eq:transformations}) into (\ref{eq:original_model}) and cancelling
$\varGamma_t\bar{u}|\bar{v}|^2$ against $-\gamma_1(t)|\bar{v}|^2\bar{u}$ we obtain
(the bars are omitted from here on)
\begin{eqnarray}\label{eq:trans_1}
iu_t+\varGamma\,u\,\big(|v|^2\big)_t+\tfrac{1}{2}\nabla^2 u
-i\varGamma\,\nabla u\!\cdot\!\nabla|v|^2
-\tfrac{i}{2}\varGamma\,u\,\nabla^2|v|^2
-\tfrac{1}{2}\varGamma^2\big|\nabla|v|^2\big|^2u
-\big(g_{11}|u|^2+\gamma_0|v|^2\big)u&=&0 , \nonumber\\
iv_t+\varGamma\,v\,\big(|u|^2\big)_t+\tfrac{1}{2}\nabla^2 v
-i\varGamma\,\nabla v\!\cdot\!\nabla|u|^2
-\tfrac{i}{2}\varGamma\,v\,\nabla^2|u|^2
-\tfrac{1}{2}\varGamma^2\big|\nabla|u|^2\big|^2v
-\big(g_{22}|v|^2+\gamma_0|u|^2\big)v&=&0 .
\end{eqnarray}
Taking the imaginary part of $u^{*}\times$(\ref{eq:trans_1}) gives the exact
balance equations
\begin{eqnarray}\label{eq:trans_2}
\big(|u|^2\big)_t&=&\frac{i}{2}\left(u^\ast\nabla^2u-u\nabla^2u^\ast\right)
+\varGamma\,\nabla|u|^2\!\cdot\!\nabla|v|^2+\varGamma\,|u|^2\nabla^2|v|^2 ,\nonumber\\
\big(|v|^2\big)_t&=&\frac{i}{2}\left(v^\ast\nabla^2v-v\nabla^2v^\ast\right)
+\varGamma\,\nabla|u|^2\!\cdot\!\nabla|v|^2+\varGamma\,|v|^2\nabla^2|u|^2 .
\end{eqnarray}
Inserting (\ref{eq:trans_2}) into (\ref{eq:trans_1}) and averaging over the
period $\Lambda=2\pi/\omega$ of the fast oscillation, using
$\langle\varGamma\rangle=0$ and
\begin{equation}\label{eq:sigma2}
\sigma^2\equiv\langle\varGamma^2(t)\rangle=\frac{\gamma_1^2}{2\omega^2},
\end{equation}
we obtain the averaged system
\begin{eqnarray}\label{eq:averaged_model}
iu_t+\tfrac{1}{2}\nabla^2 u
+\sigma^2\Big[\nabla\!\cdot\!\big(|v|^2\nabla|u|^2\big)
-\tfrac{1}{2}\big|\nabla|v|^2\big|^2\Big]u
-\big(g_{11}|u|^2+\gamma_0|v|^2\big)u&=&0 , \nonumber\\
iv_t+\tfrac{1}{2}\nabla^2 v
+\sigma^2\Big[\nabla\!\cdot\!\big(|u|^2\nabla|v|^2\big)
-\tfrac{1}{2}\big|\nabla|u|^2\big|^2\Big]v
-\big(g_{22}|v|^2+\gamma_0|u|^2\big)v&=&0 .
\end{eqnarray}
The square brackets in (\ref{eq:averaged_model}) are \emph{real}; they act as an
effective, density-dependent quantum pressure whose strength is controlled by
the population of the \emph{other} component. This is the two-dimensional
counterpart of the nonlinear dispersion found in one dimension
in~\cite{AOY_PLA}. The corresponding averaged Hamiltonian reads
\begin{equation}\label{eq:Hamiltonian_averaged}
\bar{H}=\int_{\mathbb{R}^2}\Big\{\tfrac{1}{2}|\nabla u|^2+\tfrac{1}{2}|\nabla v|^2
+\tfrac{1}{2}\sigma^2\Big(|u|^2\big|\nabla|v|^2\big|^2+|v|^2\big|\nabla|u|^2\big|^2\Big)
+\tfrac{1}{2}g_{11}|u|^4+\tfrac{1}{2}g_{22}|v|^4+\gamma_0|u|^2|v|^2\Big\}d^2x ,
\end{equation}
and $\delta\bar{H}/\delta u^\ast=iu_t$ and $\delta\bar{H}/\delta v^\ast=iv_t$ reproduces (\ref{eq:averaged_model}).

\subsection{Suppression of collapse in the averaged system}\label{sec:noblowup}

Consider the behavior of the averaged Hamiltonian~(\ref{eq:Hamiltonian_averaged}) under
the norm-preserving scaling $u_i(\mathbf{x})\to L^{-1}u_i(\mathbf{x}/L)$. One
finds~\cite{AG,Stepanov,Veks}
\begin{equation}\label{eq:HL}
\bar{H}(L)=\frac{1}{L^{2}}\sum_{i}\left(I_{1,i}+\sum_{k}g_{ik}I_{2,ik}\right)
+\frac{\sigma^2}{L^{6}}\sum_{i\neq k}I_{3,ik} ,
\end{equation}
where
\begin{eqnarray}
I_{1,i}&=&\frac{1}{2}\int_{\mathbb{R}^2}|\nabla u_i|^2 d^2 x, \qquad
I_{2,ik}=\frac{1}{2}\int_{\mathbb{R}^2}|u_i|^2|u_k|^2d^2 x, \nonumber\\
I_{3,ik}&=&\frac{1}{2}\int_{\mathbb{R}^2}|u_i|^2\big|\nabla|u_k|^2\big|^2d^2 x .
\end{eqnarray}
The crucial point is that the term generated by the management is \emph{positive
definite} and scales as $L^{-6}$, i.e.\ faster than the (possibly negative) $L^{-2}$
terms. Hence $\bar{H}(L)\to+\infty$ as $L\to0$: the energy is no longer unbounded from
below and the collapse is arrested. If the $L^{-2}$ coefficient is negative,
$\bar{H}(L)$ possesses a minimum at the finite width
\begin{equation}\label{eq:Lmin}
L_{\min}=\left(\frac{3\,\sigma^2\sum_{i\neq k}I_{3,ik}}
{\left|\sum_i\left(I_{1,i}+\sum_kg_{ik}I_{2,ik}\right)\right|}\right)^{1/4},
\end{equation}
which is precisely the NM vector soliton.

This heuristic argument can be made rigorous. Splitting
$\bar H=H_0+H_{\rm int}$ with
\begin{eqnarray}
H_0 &=& \int_{\mathbb{R}^2} \frac{1}{2}\Big(|\nabla u|^2 + |\nabla v|^2
 +\sigma^2\big[|u|^2\big|\nabla|v|^2\big|^2 \nonumber\\
 &&\qquad\qquad +|v|^2\big|\nabla|u|^2\big|^2\big]\Big)d^2 x \ \ge 0, \nonumber\\
H_{\rm int}&=&\int_{\mathbb{R}^2} \frac{1}{2}\left( g_{11}|u|^4 + g_{22}|v|^4
 + 2\gamma_0|u|^2|v|^2 \right)d^2 x ,
\end{eqnarray}
and using the Gagliardo--Nirenberg inequality in the form~\cite{AG,Stepanov}
\begin{equation}
\sum_{k,l}\|\psi_k\psi_l\|_{L^2}^2 \leq M\,C_{\rm GN}^{(2)}\,
\left(\sum_k\|\nabla\psi_k\|_{L^2}^2\right)^{3/2}
\left(\sum_k\|\psi_k\|^2_{L^2}\right)^{1/2},
\end{equation}
where $\|\cdot\|_{L^p}=(\int_{\mathbb{R}^2}|\cdot|^p d^2 x)^{1/p}$, $C_{\rm GN}^{(2)}$ is the
optimal constant of the 2D Gagliardo--Nirenberg inequality and $M=2$ is the number of
components, one shows that $|H_{\rm int}|$ grows at most like $H_0^{3/4}$ once the
$\sigma^2$ contribution to $H_0$ is taken into account, so that $\bar H$ is bounded from
below at fixed $N$. Blow-up is therefore absent in the averaged system.

We stress that this statement concerns the averaged model. The behavior of the solutions
of the full vector NLSE with time-dependent coefficients depends on the form of the
management and on the initial data, and its rigorous description remains an open
problem~\cite{Stepanov,Itin}. The direct simulations in Sec.~\ref{sec_Numerical_results} show,
however, that over the times explored here the full model follows the averaged one very
closely.

\section{Variational approach}\label{sec_3}

\subsection{Effective Lagrangian and equations of motion}

The averaged system (\ref{eq:averaged_model}) follows from the Lagrangian
density
\begin{eqnarray}\label{eq:LD}
L&=&\frac{i}{2}\left(u\,u^\ast_t - u^\ast u_t\right)
+\frac{i}{2}\left(v\,v^\ast_t - v^\ast v_t\right)
+\frac{1}{2}\left|\nabla u\right|^2+\frac{1}{2}\left|\nabla v\right|^2
\nonumber \\
&&+\frac{1}{2}g_{11}|u|^4+\frac{1}{2}g_{22}|v|^4+\gamma_0|u|^2|v|^2
+\frac{\sigma^2}{2}\Big(|u|^2\big|\nabla|v|^2\big|^2+|v|^2\big|\nabla|u|^2\big|^2\Big).
\end{eqnarray}
We use radially symmetric Gaussian trial functions with real time-dependent
amplitudes $A_i(t)$, widths $w_i(t)$, chirps $b_i(t)$ and phases $\phi_i(t)$,
\begin{equation}\label{eq:ansatz}
u=A_1e^{-\rho^2/(2w_1^2)+ib_1\rho^2+i\phi_1},\qquad
v=A_2e^{-\rho^2/(2w_2^2)+ib_2\rho^2+i\phi_2},\qquad \rho^2=x^2+y^2 ,
\end{equation}
and impose the variational principle $\delta\int\bar{L}\,dt=0$ on the reduced
Lagrangian $\bar{L}=\int_{\mathbb{R}^2}L\,d^2x$~\citep{Anderson}. Carrying out the integrations
gives, with $N_u=\pi A_1^2w_1^2$ and $N_v=\pi A_2^2w_2^2$,
\begin{eqnarray}\label{eq:Lbar}
\bar{L}&=&N_u\Big(w_1^2\dot{b}_1+\dot{\phi}_1+2b_1^2w_1^2\Big)
       +N_v\Big(w_2^2\dot{b}_2+\dot{\phi}_2+2b_2^2w_2^2\Big)
\nonumber\\
&&+\frac{N_u}{2w_1^2}+\frac{N_v}{2w_2^2}
+\frac{g_{11}N_u^2}{4\pi w_1^2}+\frac{g_{22}N_v^2}{4\pi w_2^2}
+\frac{\gamma_0 N_uN_v}{\pi\left(w_1^2+w_2^2\right)}
\nonumber\\
&&+\frac{2\sigma^2}{\pi^2}\left[
\frac{N_u^2N_v\,w_2^2}{w_1^4\left(w_1^2+2w_2^2\right)^2}
+\frac{N_uN_v^2\,w_1^2}{w_2^4\left(2w_1^2+w_2^2\right)^2}\right].
\end{eqnarray}
Variation with respect to $\phi_i$ gives $N_u,N_v=$~const, variation with
respect to $b_i$ gives $\dot{w_i}=2b_iw_i$, and variation with respect to
$w_i$ then yields the following closed system for the widths
\begin{eqnarray}
\frac{d^2w_1}{dt^2}&=& \frac{1}{w_{1}^{3}} + \frac{g_{11} N_{u}}{2\pi w_{1}^{3}}
+ \frac{2\gamma_{0}N_{v} w_{1}}{\pi \left(w_{1}^{2} + w_{2}^{2}\right)^2}
+\frac{4\sigma^2}{\pi^2}\left[\frac{4 N_{u} N_{v} w_{2}^{2}}
{\left(w_{1}^{2} + 2 w_{2}^{2}\right)^{3} w_{1}^{3}}
\left(\frac{w_2^2}{w_1^2}+1\right)
+ \frac{N_{v}^{2} w_{1}}{\left(2 w_{1}^{2} + w_{2}^{2}\right)^{3} w_{2}^{2}}
\left(\frac{2w_1^2}{w_2^2}-1\right)\right] , \label{eq:w1_tt}\\
\frac{d^2w_2}{dt^2}&=& \frac{1}{w_{2}^{3}} + \frac{g_{22} N_{v}}{2\pi w_{2}^{3}}
+ \frac{2\gamma_{0}N_{u} w_{2}}{\pi \left(w_{1}^{2} + w_{2}^{2}\right)^2}
+\frac{4\sigma^2}{\pi^2}\left[\frac{4 N_{u} N_{v} w_{1}^{2}}
{\left(2 w_{1}^{2} + w_{2}^{2}\right)^{3} w_{2}^{3}}
\left(\frac{w_1^2}{w_2^2}+1\right)
+ \frac{N_{u}^{2} w_{2}}{\left(w_{1}^{2} + 2w_{2}^{2}\right)^{3} w_{1}^{2}}
\left(\frac{2w_2^2}{w_1^2}-1\right)\right] . \label{eq:w2_tt}
\end{eqnarray}
It is convenient to notice that (\ref{eq:w1_tt})--(\ref{eq:w2_tt}) describe a
two-dimensional mechanical particle with anisotropic masses $(N_u,N_v)$,
\begin{equation}\label{eq:mechanical}
N_u\ddot{w}_1=-\frac{\partial U}{\partial w_1},\qquad
N_v\ddot{w}_2=-\frac{\partial U}{\partial w_2},\qquad
E=\frac{1}{2}N_u\dot{w}_1^2+\frac{1}{2}N_v\dot{w}_2^2+U(w_1,w_2)=\bar{H} ,
\end{equation}
moving in the effective potential
\begin{equation}\label{eq:Upot}
U(w_1,w_2)=\frac{N_u}{2w_1^2}+\frac{N_v}{2w_2^2}
+\frac{g_{11}N_u^2}{4\pi w_1^2}+\frac{g_{22}N_v^2}{4\pi w_2^2}
+\frac{\gamma_0N_uN_v}{\pi(w_1^2+w_2^2)}
+\frac{2\sigma^2}{\pi^2}\left[\frac{N_u^2N_vw_2^2}{w_1^4(w_1^2+2w_2^2)^2}
+\frac{N_uN_v^2w_1^2}{w_2^4(2w_1^2+w_2^2)^2}\right],
\end{equation}
whose value on a trajectory coincides with the averaged Hamiltonian
(\ref{eq:Hamiltonian_averaged}) evaluated on the ansatz. For $\sigma=0$ the
potential is a pure $-|\mathcal{A}|/w^2$ well with no minimum, and the particle
reaches $w=0$ in finite time (collapse); the $\sigma^2$ terms add a
$w^{-6}$ repulsive core which is precisely the mechanism of collapse arrest.

An equivalent derivation, which does not rely on the averaged
equations, uses the modified variational ansatz of~\cite{AG} in which the
rapid cross-phase modulation is built into the trial function,
\begin{eqnarray}\label{eq:ansatz1}
u&=&A_1e^{-\rho^2/(2w_1^2)+ib_1\rho^2+i\phi_1}
\exp\!\big[-i\varGamma(t)A_2^2e^{-\rho^2/w_2^2}\big] ,\nonumber\\
v&=&A_2e^{-\rho^2/(2w_2^2)+ib_2\rho^2+i\phi_2}
\exp\!\big[-i\varGamma(t)A_1^2e^{-\rho^2/w_1^2}\big] .
\end{eqnarray}
Substituting (\ref{eq:ansatz1}) into the Lagrangian of the \emph{original}
model and averaging over the period of the fast modulation reproduces
(\ref{eq:Lbar}) exactly. We stress that the fast phase in (\ref{eq:ansatz1})
is essential: the plain chirped Gaussian~(\ref{eq:ansatz}) applied directly to
the unaveraged model, followed by a Kapitza-type averaging of the resulting
Mathieu-like width equation, produces a stabilising term
$3\sigma^2N^2/(4\pi^2w^7)$ instead of $4\sigma^2N^2/(3\pi^2w^7)$, i.e. it
underestimates the effective pressure by the factor $9/16$. The chirped
Gaussian simply cannot represent the phase $\exp[-i\varGamma|v|^2]$ that the
strong management imprints on the field, and the averaged
system~(\ref{eq:averaged_model}) must be used.

\subsection{Symmetric solitons}\label{sec:symmetric}

For $g_{11}=g_{22}=g$, $N_u=N_v=N_0$ and $w_1=w_2=w$, Eqs.
(\ref{eq:w1_tt})--(\ref{eq:w2_tt}) collapse onto the single equation
\begin{equation}\label{eq:width}
\frac{d^2w}{dt^2}=\frac{\alpha}{w^3}+\frac{4\sigma^2N_0^2}{3\pi^2 w^7},
\qquad
\alpha\equiv 1+\frac{(g+\gamma_0)N_0}{2\pi}=1-\frac{N_0}{N_c},
\end{equation}
with the mechanical potential (per unit mass, $U_{\rm sym}=2N_0\,\mathcal{U}$)
\begin{equation}\label{eq:Usym}
\mathcal{U}(w)=\frac{\alpha}{2w^2}+\frac{2\sigma^2N_0^2}{9\pi^2w^6} .
\end{equation}
Note that $\alpha$ is exactly the collapse discriminant of
Sec.~\ref{sec:virial}: $\alpha<0$ ($N_0>N_c$) is the parameter region in which the unmodulated condensate blows up. A stationary state exists precisely in that region, and its width follows from $\mathcal{U}'(w_0)=0$,
\begin{equation}\label{eq:w0}
w_0=\left(-\frac{4\sigma^2N_0^2}{3\pi^2\alpha}\right)^{1/4}
=\left(\frac{8\sigma^2N_0^2}{3\pi\left[-2\pi-(g+\gamma_0)N_0\right]}\right)^{1/4},
\qquad
A_0=\sqrt{\frac{N_0}{\pi w_0^2}} .
\end{equation}
Thus \emph{nonlinearity management converts the collapse region of the
unmanaged system into the existence region of a stable localised state}.
The depth of the well and the corresponding averaged energy are
\begin{equation}\label{eq:binding}
\mathcal{U}(w_0)=\frac{\alpha}{3w_0^2}<0,\qquad
\bar{H}_{\min}=\frac{2N_0\alpha}{3w_0^2} ,
\end{equation}
and, since $\mathcal{U}(w\to\infty)\to 0^-$, the soliton is dynamically bound only as long as $\bar{H}<0$; $\bar{H}>0$ leads to unlimited expansion. This gives a sharp criterion for the largest admissible perturbation of the stationary state.

\paragraph{Normal modes.} Linearising (\ref{eq:w1_tt})--(\ref{eq:w2_tt}) about
$w_1=w_2=w_0$ we find two eigenmodes. The in-phase (breathing) mode
$\delta w_1=\delta w_2$ and the out-of-phase (symmetry-breaking) mode
$\delta w_1=-\delta w_2$ oscillate with
\begin{equation}\label{eq:Omegas}
\Omega_-=\frac{2\sqrt{-\alpha}}{w_0^2},\qquad
\Omega_+=\frac{2\sqrt{-\alpha_+}}{3\,w_0^2},\qquad
\alpha_+\equiv 1+\frac{(g+10\gamma_0)N_0}{2\pi}=\alpha+\frac{9\gamma_0N_0}{2\pi} .
\end{equation}
The breathing frequency $\Omega_-$ can equivalently be written as
$\Omega_-^2=\mathcal{U}''(w_0)=[3\alpha+28\sigma^2N_0^2/(3\pi^2w_0^4)]/w_0^4$, and,
on eliminating $w_0$ with (\ref{eq:w0}), in the compact invariant form
\begin{equation}\label{eq:Omcompact}
\Omega_-=\frac{\pi\sqrt{3}\,|\alpha|}{\sigma N_0}
=\frac{\sqrt{3}\,\big|2\pi+(g+\gamma_0)N_0\big|}{2\,\sigma N_0} ,
\end{equation}
which displays the scaling $\Omega\propto(\sigma N_0)^{-1}$ explicitly: the
breathing frequency is set by the distance $|\alpha|$ from the collapse
threshold divided by the management strength $\sigma N_0$, so a stronger
management produces a wider and more slowly breathing soliton,
$w_0\propto(\sigma N_0)^{1/2}$.
Since existence requires $\alpha<0$, the breathing mode is always stable. 

\begin{center}
\begin{figure}[H]
\includegraphics[scale=0.7]{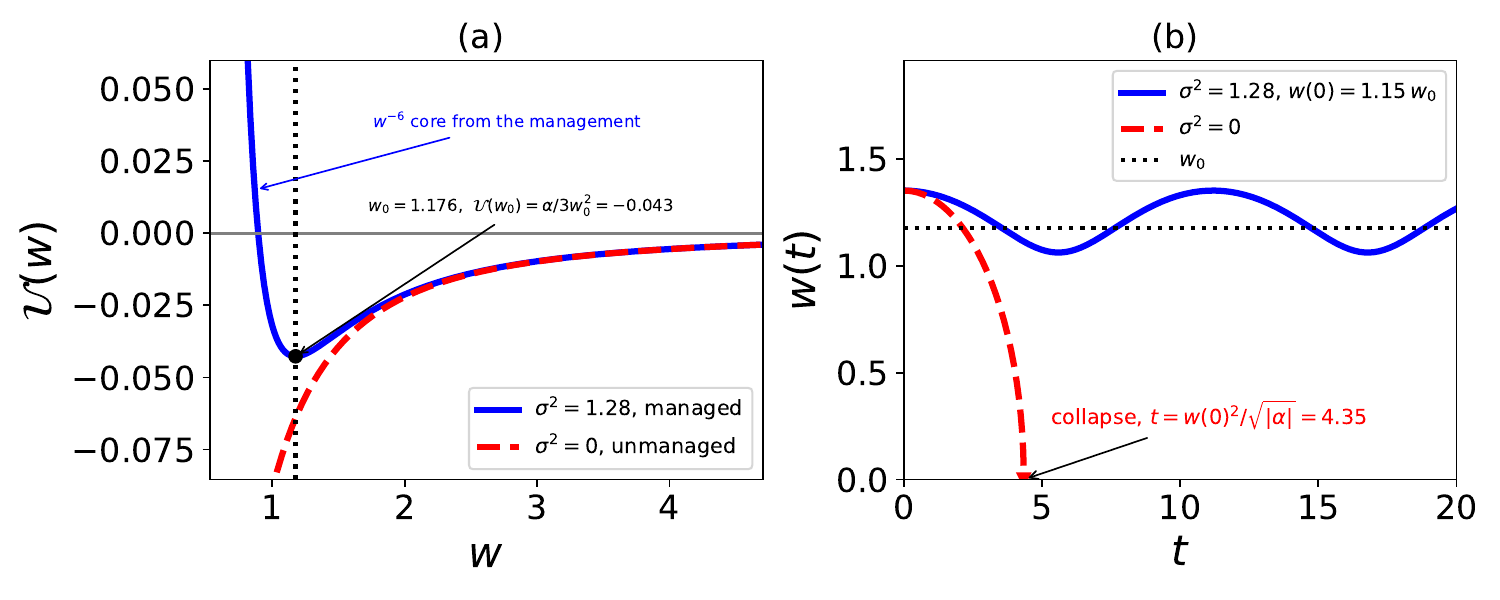}
\caption{(Color online) (a) Effective potential $\mathcal{U}(w)$,
Eq.~(\ref{eq:Usym}), for the parameters in Set B~(table \ref{tab:sets}). Without management ($\sigma^2=0$, red dashed) the potential is attractive at all scales and the width reaches zero in a finite time; the nonlinear quantum pressure created by the modulation (blue solid) adds a repulsive $w^{-6}$ core and produces a minimum at $w_0$. (b) Width $w(t)$ from the variational equation (\ref{eq:width}) for the same parameters and the initial condition $w(0)=1.15\,w_0$, $\dot w(0)=0$: collapse for $\sigma^2=0$ (red dashed) and bounded oscillations with frequency $\Omega_-$ for $\sigma^2\neq0$ (blue solid).}
\label{fig:1}
\end{figure}
\end{center}

Figure~\ref{fig:1} presents a comparison between the unmanaged (\(\sigma^2=0\)) and managed (\(\sigma^2\neq0\)) scenarios using the parameters from Set $B$ (table~\ref{tab:sets}). As shown in the effective potential curves (Fig.~\ref{fig:1}(a)), the time-modulation injects a repulsive \(w^{-6}\) core that counteracts the pure attraction of the unmanaged system. Consequently, instead of a catastrophic collapse where the width reaches zero, the variational dynamics (Fig.~\ref{fig:1}(b)) demonstrate that the nonlinear quantum pressure successfully stabilizes the condensate, leading to regular bounded oscillations with frequency \(\Omega _{-}\).

\subsection{Townes-corrected estimate}\label{sec:townes}

The accuracy of (\ref{eq:w0}) is limited by a single, well-understood defect of
the Gaussian ansatz. In the symmetric sector the averaged system reduces to a
scalar equation with effective coupling $g+\gamma_0$, whose unmanaged
($\sigma=0$) critical norm is set by the Townes mode $R$~\citep{Vlasov},
\begin{equation}\label{eq:NcT}
N_c^{\rm ex}=\frac{\|R\|^2_{L^2}}{2|g+\gamma_0|}\simeq\frac{5.850}{|g+\gamma_0|},
\qquad\text{whereas}\qquad
N_c^{\rm G}=\frac{2\pi}{|g+\gamma_0|},
\end{equation}
so the Gaussian trial function overestimates $N_c$ by the known factor
$2\pi/5.850=1.074$. Since $\alpha=1-N_0/N_c$ is precisely this collapse
discriminant, replacing $N_c^{\rm G}$ by $N_c^{\rm ex}$ in
$\alpha$ --- while keeping the Gaussian coefficient of the nonlinear-pressure
term, which is far less sensitive to the profile --- gives the corrected width
\begin{equation}\label{eq:w0corr}
w_0^{\rm corr}=\left(-\frac{4\sigma^2N_0^2}{3\pi^2\alpha_{\rm ex}}\right)^{1/4},
\qquad \alpha_{\rm ex}=1-\frac{N_0}{N_c^{\rm ex}} .
\end{equation}
As shown in Sec.~\ref{sec_Numerical_results}, (\ref{eq:w0corr}) reproduces the
numerically exact stationary widths of the averaged system to better than $1\%$
for $N_0\gtrsim1.2N_c^{\rm ex}$, against an error of $9$--$40\%$ for the
uncorrected formula (\ref{eq:w0}). 

\subsection{Scaling invariances}\label{sec:pdescaling}\label{sec:scaling}
 
Two exact invariances organise all the results below. They belong to the
averaged system~(\ref{eq:averaged_model}) itself and not to the Gaussian
ansatz; the reduced equations (\ref{eq:w1_tt})--(\ref{eq:w2_tt}) merely inherit
them, which is a useful check on the algebra of the preceding subsections.
 
\emph{(i) Norm scaling.} Writing $u=\sqrt{N}\,\tilde{u}$,
$v=\sqrt{N}\,\tilde{v}$ with $\int|\tilde{u}|^2d^2x=\int|\tilde{v}|^2d^2x=1$,
the averaged system depends on the parameters only through the combinations
\begin{equation}\label{eq:normscaling}
g_{ij}N \quad\text{and}\quad \sigma N ,
\end{equation}
and the original system~(\ref{eq:original_model}) only through $g_{ij}N$,
$\gamma_1N$ and $\omega$. Absolute values of $N$ and of the couplings are
therefore meaningless on their own, and every result below is quoted in terms
of these invariant combinations; for the two-component problem the complete
list is $g_{11}N_u$, $g_{22}N_v$, $\gamma_0N$ and $N_v/N_u$, with $\sigma N$
fixing the scale.
 
\emph{(ii) Length scaling.} The averaged system is invariant under
\begin{equation}\label{eq:lenscaling}
u\to\lambda^{-1}u(\mathbf{x}/\lambda,\,t/\lambda^2),\qquad
v\to\lambda^{-1}v(\mathbf{x}/\lambda,\,t/\lambda^2),\qquad
\sigma^2\to\lambda^{4}\sigma^2 ,
\end{equation}
at fixed $N_u$, $N_v$ and $g_{ij}$, because the kinetic and mean-field terms
both scale as $\lambda^{-3}$ while the nonlinear-pressure terms scale as
$\lambda^{-7}$; on the widths it reads $w_i\to\lambda w_i$, $t\to\lambda^2t$.
The management therefore does not change the \emph{shape} of the family of
localised states, only its scale:
\begin{equation}\label{eq:scaling}
w_i\propto(\sigma N)^{1/2}=2^{-1/4}\left(\frac{\gamma_1N}{\omega}\right)^{1/2},
\qquad
\Omega_-\, (\Omega_+)\propto(\sigma N)^{-1} = \frac{\pi\sqrt{3}\,|\alpha|}{\sigma N_0} \quad \left(\frac{\pi\,\sqrt{\alpha \alpha_+}}{\sqrt{3} \sigma N_0} \right).
\end{equation}
 
A useful corollary follows by evaluating the strength of the nonlinear
pressure at the fixed point: with $A_0^2=N_0/(\pi w_0^2)$ one finds from
(\ref{eq:w0}) the universal relation
\begin{equation}\label{eq:universal}
4\sigma^2A_0^4=-3\alpha=3\left(\frac{N_0}{N_c}-1\right).
\end{equation}
The dimensionless parameter $4\sigma^2A_0^4$ is precisely the correction to the
$k^4$ dispersion coefficient of the averaged system; Eq.~(\ref{eq:universal})
therefore states that the averaged description is asymptotically controlled
only when the norm does not exceed the critical one too strongly. This
explains why, in the simulations of Sec.~\ref{sec_Numerical_results}, NM vector
solitons are long-lived close above $N_c$ and rapidly destroyed for
$N_0\gg N_c$.

\subsection{Asymmetric solitons}\label{sec:asymmetric}
 
For $N_u\neq N_v$ the stationary widths are the roots of the coupled algebraic
system obtained by setting the right-hand sides of
(\ref{eq:w1_tt})--(\ref{eq:w2_tt}) to zero, equivalently the critical points of
$U(w_1,w_2)$ in (\ref{eq:Upot}). They are found by Newton iteration, continued
in $N_v$ at fixed $N_u$ from the symmetric solution (\ref{eq:w0}) --- or, when
the two components are very unequal, from the seed
$(w,\kappa_*w)$ with $\kappa_*$ given below. Their linear stability is decided
by the eigenvalues of $M^{-1}K$ with $M={\rm diag}(N_u,N_v)$ and
$K_{ij}=\partial^2U/\partial w_i\partial w_j$ evaluated at the fixed point; the
state is stable when both eigenvalues are positive, i.e.\ when the critical
point is a minimum of $U$.
 
The existence boundary of this family can be obtained in closed form. For large
widths the $\sigma^2$ terms in (\ref{eq:Upot}) decay as $w^{-6}$ and
\begin{equation}
U(w_1,w_2)\ \longrightarrow\ \frac{a_1}{w_1^2}+\frac{a_2}{w_2^2}
+\frac{a_{12}}{w_1^2+w_2^2},
\end{equation}
which is exactly the Gaussian energy (\ref{eq:Hw1w2}). A bound stationary state
therefore exists if and only if this asymptotic form is negative somewhere, i.e.\ precisely under condition (\ref{eq:sharpNc}):
\begin{equation}\label{eq:asymbound}
\text{existence}\ \Longleftrightarrow\
\sqrt{2\pi N_u+g_{11}N_u^2}+\sqrt{2\pi N_v+g_{22}N_v^2}<2\sqrt{-\gamma_0N_uN_v} .
\end{equation}
Where (\ref{eq:asymbound}) becomes an equality the optimal ratio is
$w_2/w_1\to\kappa_*=(a_2/a_1)^{1/4}$, both widths diverge as $U$ flattens, and
both normal-mode frequencies vanish.
 
\paragraph{A second limit.}
Equation (\ref{eq:asymbound}) is not the only way to destroy the soliton, and
it is not even the binding one when an intra-species interaction is attractive.
The nonlinear pressure acting on one component is proportional to the density
of the \emph{other}, so it switches off for a component left on its own; a
component must therefore stay below its own Townes threshold,
\begin{equation}\label{eq:ceiling}
g_{ii}N_i>-2\pi ,
\end{equation}
or it can collapse alone with nothing to arrest it. For $g_{ii}>0$ the
condition (\ref{eq:ceiling}) is vacuous and (\ref{eq:asymbound}) closes the
family at both ends; for $g_{ii}<0$ it can bind first, and then the family
simply runs out of validity while $U$ still has a perfectly good minimum ---
something $U(w_1,w_2)$ alone cannot see. The two parameter sets of
Sec.~\ref{sec_Numerical_results} realise the two cases.
 
\begin{figure}[H]
\includegraphics[scale=0.6]{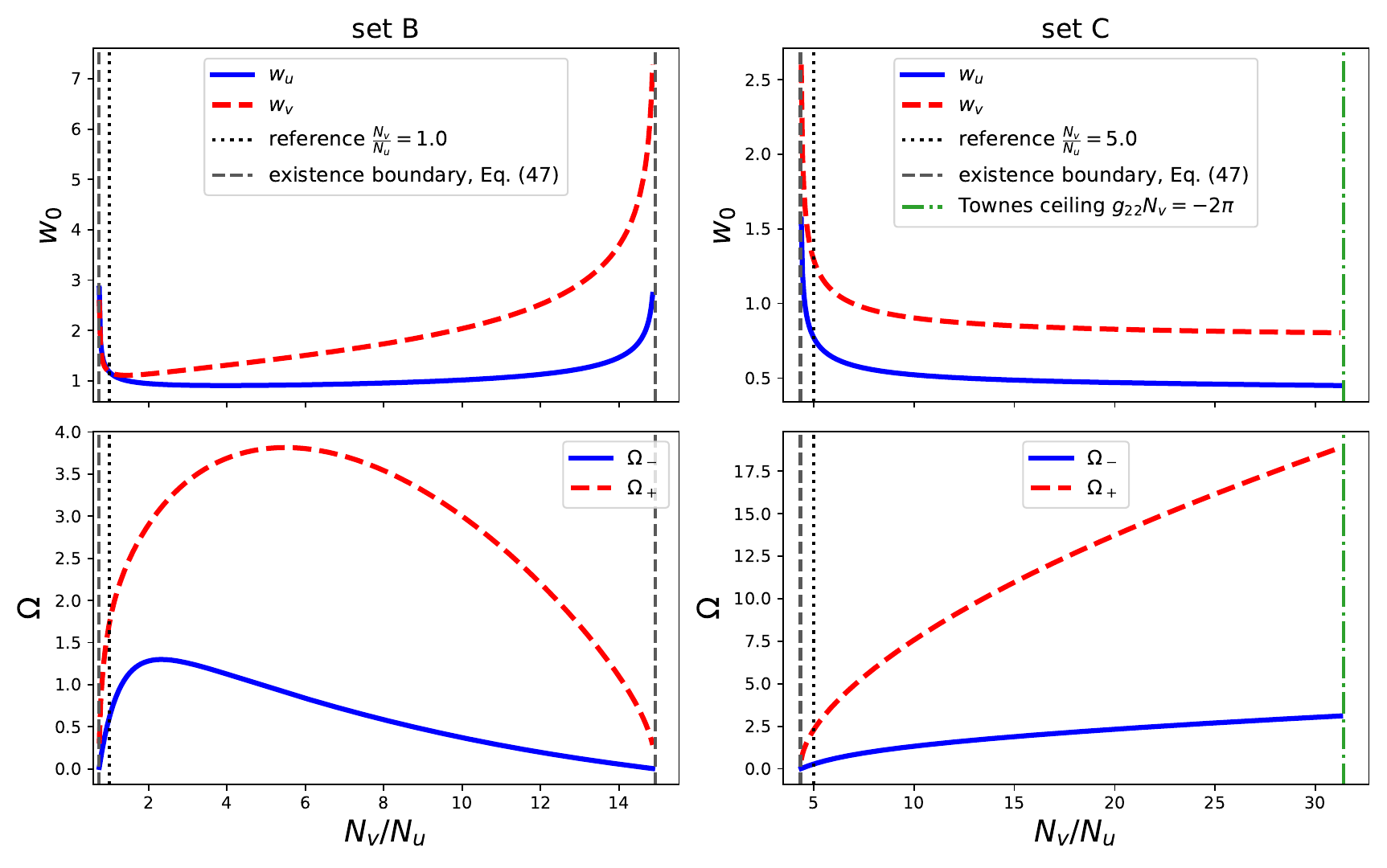}
\caption{(Color online) Asymmetric NM vector solitons, by Newton continuation
of the fixed points of $U(w_1,w_2)$ in $N_v$ at fixed $N_u$. Left: set~B ($g_{11}=g_{22}=1$, $\gamma_0=-2\pi$, $N_u=1.4$); right: set~C ($g_{11}=-4$, $g_{22}=-1/4$, $\gamma_0=-\pi$, $N_u=0.8$). (Upper two profile) stationary widths $w_u$ (blue) and $w_v$ (red dashed); (Lower two profile) the normal-mode frequencies, both real over the whole family. Grey dashed lines are the existence boundaries
(\ref{eq:asymbound}); the green dash-dotted line in the right column is the single-component Townes ceiling (\ref{eq:ceiling}), which for set~C is reached before the existence boundary and terminates the family while $U$ still has a minimum. Dotted black: the reference imbalance used in Sec.~\ref{sec_Numerical_results}.}
\label{fig:2}
\end{figure}

\paragraph{Set B ($g_{11}=g_{22}=1$, $\gamma_0=-2\pi$, $\gamma_1=8\pi$,
$\omega=5\pi$, $N_u=1.4$).}
Both intra-species interactions are repulsive, so $a_1,a_2>0$ for every $N_v$
and (\ref{eq:asymbound}) decides the whole family. It gives
\begin{equation}
1.0330<N_v<20.9078 ,\qquad\text{i.e.}\qquad 0.738<\frac{N_v}{N_u}<14.934 ,
\end{equation}
with $\kappa_*=0.916$ at the lower edge and $\kappa_*=2.696$ at the upper one.
Figure~\ref{fig:2} (Set B in the table~\ref{tab:sets}) shows the continuation: the widths diverge and
$\Omega_-\to0$ at \emph{both} ends, exactly as the closed form predicts, while
in between the family is a stable one-parameter branch
($\min\Omega_-=4.4\times10^{-3}$ over the whole range, attained only at the
edges). At the reference point $N_v=N_u$ it reduces to the symmetric soliton,
$w_u=w_v=1.176$, $\Omega_-=0.6087$, and $\Omega_+=1.7230$. Population imbalance
therefore destroys the managed soliton in the same way as an insufficient norm,
and a single inequality controls both.
 
\paragraph{Set C ($g_{11}=-4$, $g_{22}=-1/4$, $\gamma_0=-\pi$,
$\gamma_1=6\pi$, $\omega=40$, $N_u=0.8$).}
Here all three interactions are attractive and the picture changes at the upper
end. The lower boundary is still given by (\ref{eq:asymbound}),
\begin{equation}
N_v>3.4677 ,\qquad\text{i.e.}\qquad \frac{N_v}{N_u}>4.335 ,
\qquad \kappa_*=1.661 ,
\end{equation}
and the widths diverge there in the usual way. Above it, however, the left-hand
side of (\ref{eq:asymbound}) never catches up again: $a_2=(2\pi
N_v+g_{22}N_v^2)/4\pi$ \emph{decreases} once $N_v>\pi/|g_{22}|$ and reaches zero
at the ceiling (\ref{eq:ceiling}),
\begin{equation}
N_v<\frac{2\pi}{|g_{22}|}=25.133 ,\qquad\text{i.e.}\qquad
\frac{N_v}{N_u}<31.416 ,
\end{equation}
which is reached first. Figure~\ref{fig:2} (Set C in the table~\ref{tab:sets}) shows the consequence: the
variational family neither flattens nor loses stability as $N_v$ grows --- the
widths saturate, $w_u\to0.45$ and $w_v\to0.80$, and both frequencies increase
monotonically --- and it is terminated by a mechanism that the two-width
potential does not contain. At the reference imbalance $N_v/N_u=5.0$ used in
Sec.~\ref{sec_Numerical_results} the soliton is comfortably inside both limits,
\begin{equation}
w_u=0.768,\quad w_v=1.29,\quad \frac{w_v}{w_u}\approx 1.677\ (\kappa_*=1.787),
\qquad \Omega_-=0.2844,\quad \Omega_+=2.3 ,
\end{equation}
with amplitudes $A_u=0.656$ and $A_v=0.874$: the two components differ by
$68\%$ in width and $33\%$ in amplitude, so the state is a genuine vector
soliton and not a single-component one dressed by a passive partner.
 
Comparing the two sets makes the design rule clear. What drives the contrast between the two components is not the population imbalance, but the ratio $\kappa_*=a_2/a_1$. In set~B, for instance, even at $N_v/N_u=1.5$, $\kappa_*$ stays as low as $1.13$ (with $w_v/w_u=1.12$), while set~C reaches $\kappa_*=1.79$ simply because $g_{11}N_u=-3.20$ and $g_{22}N_v=-1.25$ are far apart. In other words, an asymmetric NM vector soliton comes from separating $g_{11}N_u$ from $g_{22}N_v$, not from loading more population into one component.
 
\section{Numerical results}\label{sec_Numerical_results}

\subsection{Methods and parameters}\label{sec:methods}

The stationary states of the averaged system are obtained by the \emph{dynamical functional particle method} (DFPM) with damping parameter $\eta = 2$~\cite{Magnus_DFPM_1, Magnus_DFPM_2}. Meanwhile, both PDEs are integrated using the \emph{same} second-order Strang split-step Fourier scheme,
\begin{equation}\label{eq:strang}
\mathcal{A}(\tfrac{\Delta t}{2})\;
\mathcal{N}(\tfrac{\Delta t}{2})\;
\mathcal{K}(\Delta t)\;
\mathcal{N}(\tfrac{\Delta t}{2})\;
\mathcal{A}(\tfrac{\Delta t}{2}),
\end{equation}
where $\mathcal{K}$ is the kinetic propagator $e^{-\frac{i}{2}k^2\Delta t}$, $\mathcal{A}$ the absorber $e^{-W\Delta t/2}$ and $\mathcal{N}$ the derivative-free part. Each of the three substeps is solved \emph{exactly}, so the scheme is unconditionally stable and $\Delta t$ is limited by accuracy alone.

\paragraph{Grid and step.}
All runs use the square $[-40,40]^2$ with $1024\times1024$ points,
$\Delta x = \Delta y = 0.0781$. The time step is $\Delta t=10^{-3}$ for the averaged model,
which has no fast scale, and $\Delta t$ is such that there are about $800$ steps
per modulation period for the original model
($5\times10^{-4}$ for $\omega=5\pi$, $2\times10^{-4}$ for $\omega=40$).

\paragraph{Parameters.}\label{sec:params}
We use three parameter sets, listed in Table~\ref{tab:sets}. Sets~A and~B are
symmetric ($N_u=N_v\equiv N_0$, $g_{11}=g_{22}\equiv g$) and are chosen to sit
at practically the same distance above the exact critical norm,
$N_0/N_c^{\rm ex}=1.28$ and $1.26$, while differing in everything else: in
set~A the intra-species interaction is \emph{attractive} and the management is
weak ($\sigma^2=0.08$), in set~B it is \emph{repulsive} and the management is
strong ($\sigma^2=1.28$). Since only the products $g_{ij}N_i$, $\gamma_0N$ and
$\sigma N$ are meaningful (Sec.~\ref{sec:pdescaling}), these are two genuinely
different physical points and not one point in disguise: $gN_0=-3.75$ against
$+1.40$, $\sigma N_0=2.12$ against $1.58$. Set~C is asymmetric, with all three
interactions attractive and with $N_u\ne N_v$; it is analysed in
Sec.~\ref{sec:asymnum}.

\begin{table}[H]
\caption{The three parameter sets. $\sigma^2=\gamma_1^2/(2\omega^2)$;
$N_c^{\rm ex}=\|R\|_{L^2}^2/(2|g+\gamma_0|)$;
$\alpha_{\rm ex}=1-N_0/N_c^{\rm ex}$. The last two lines are the scale-invariant
combinations that actually determine the physics.}
\label{tab:sets}
\begin{ruledtabular}
\small
\begin{tabular}{lccccccc}
 & $g_{11}$ & $g_{22}$ & $\gamma_0$ & $\gamma_1$ & $\omega$ & $N_u$ & $N_v$ \\ \hline
A & $-0.5$  & $-0.5$  & $-0.5$   & $2\pi$ & $5\pi$  & $7.5$ & $7.5$ \\
B & $1$     & $1$     & $-2\pi$  & $8\pi$ & $5\pi$  & $1.4$ & $1.4$ \\
C & $-4$    & $-1/4$  & $-\pi$   & $6\pi$ & $40$    & $0.8$ & $4$   \\
\hline \hline
 & $\sigma^2$ & $g+\gamma_0$ & $N_c^{\rm ex}$ & $N_0/N_c^{\rm ex}$ & $\alpha$ & $\alpha_{\rm ex}$ & $\omega/\Omega_-$ \\ \hline
A & $0.08$   & $-1$      & $5.8504$ & $1.2820$ & $-0.1937$ & $-0.2820$ & $31.6$ \\
B & $1.28$   & $-5.2832$ & $1.1074$ & $1.2643$ & $-0.1772$ & $-0.2643$ & $25.8$ \\
C & $0.1110$ & ---       & ---      & ---      & ---       & ---       & $17.4$  \\
\hline \hline
\multicolumn{8}{l}{$g_{11}N_u$, $g_{22}N_v$ \qquad\qquad\quad\, \textbf{A:} $-3.75$ \qquad\quad \textbf{B:} $+1.40$ \qquad\quad \textbf{C:} $-3.20$, $-1.0$} \\
\multicolumn{8}{l}{$\gamma_0\sqrt{N_uN_v}$, $\sigma\sqrt{N_uN_v}$ \quad \textbf{A:} $-3.75$, $2.12$ \quad \textbf{B:} $-8.80$, $1.58$ \quad \textbf{C:} $-5.62$, $0.6$}
\end{tabular}
\end{ruledtabular}
\end{table}

\subsection{Stationary states of the averaged system}\label{sec:statnum}

\begin{center}
\begin{figure}[H]
\includegraphics[scale=0.6]{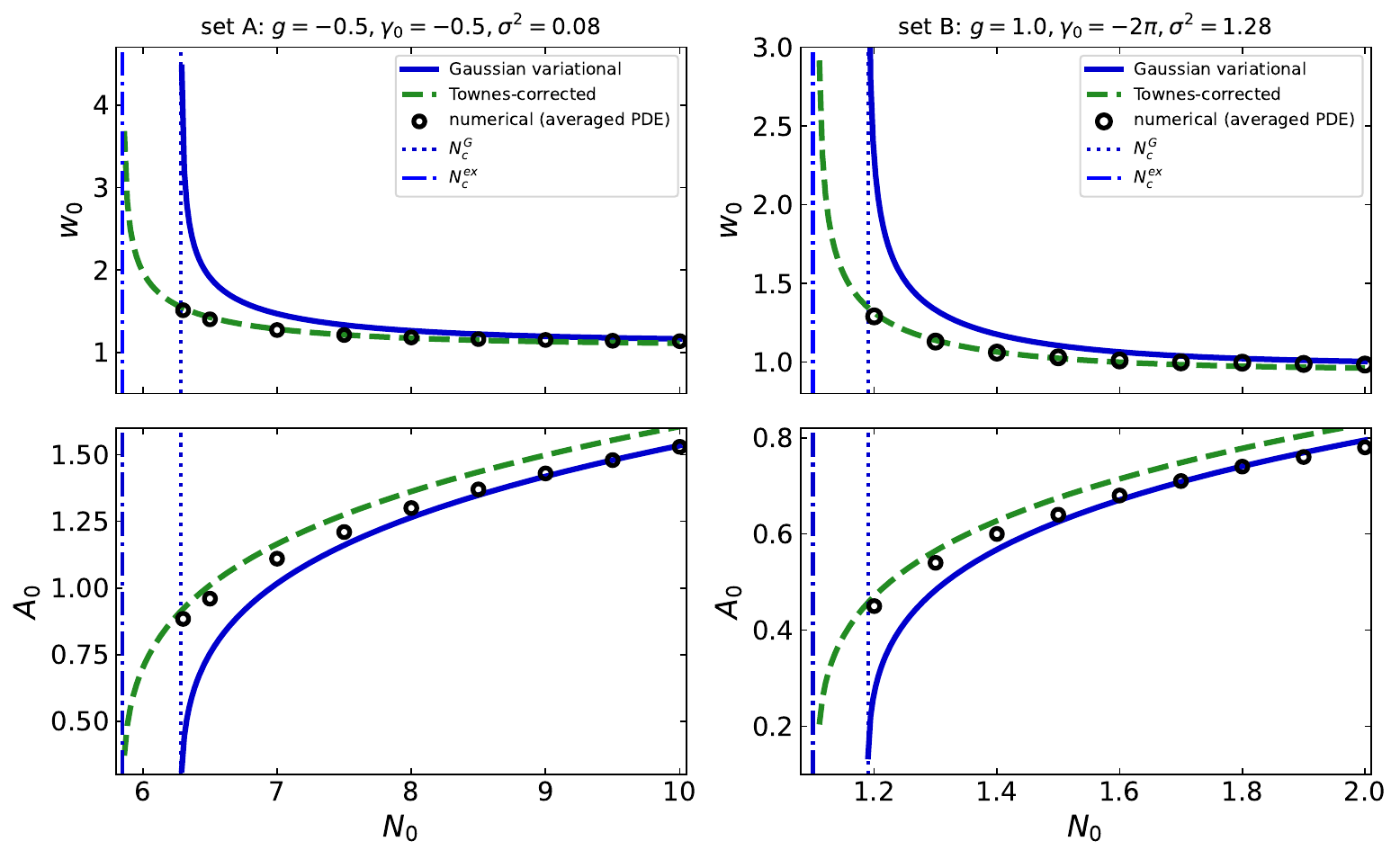}
\caption{(Color online) Width (top) and amplitude (bottom) of the stationary
NM vector soliton versus the norm per component, for set~A (left) and set~B(right). Circles: DFPM states of the averaged system; solid: Gaussian variational prediction (\ref{eq:w0}); dashed: Townes-corrected
(\ref{eq:w0corr}). Dotted and dashed-dotted vertical lines mark $N_c^{\rm G}$ and
$N_c^{\rm ex}$, respectively.}
\label{fig:3}
\end{figure}
\end{center}

Figure~\ref{fig:3} shows the width and the amplitude of the symmetric
stationary state as functions of the norm per component, for sets~A and~B,
together with the Gaussian prediction (\ref{eq:w0}) and its Townes-corrected
form (\ref{eq:w0corr}). The picture is the same in both sets although their
invariants differ by a factor of three, which is the scaling
invariance~(\ref{eq:normscaling}) at work: the uncorrected Gaussian
overestimates the width by $9$--$13\,\%$ near the threshold, while replacing
$N_c^{\rm G}$ by $N_c^{\rm ex}$ brings the prediction to within $1\,\%$ of the
DFPM state for $N_0\gtrsim1.2\,N_c^{\rm ex}$. At the reference norm of set~B
the agreement is remarkable: $w_0^{\rm corr}=1.06418$ against the DFPM value
$1.06472$, a relative error of $5\times10^{-4}$, with
$A_0=0.59986$, $\mu=-0.43418$, and $\bar H=-0.16391$.

\subsection{Arrest of the collapse}\label{sec:arrest}

\begin{center}
\begin{figure}[H]
\includegraphics[scale=0.35]{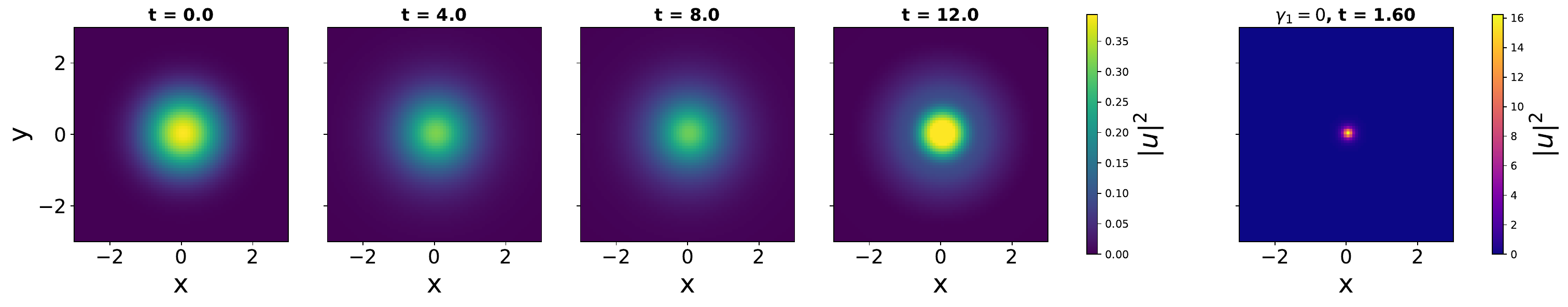}
\caption{(Color online) Density $|u|^2$ of the full model
(\ref{eq:original_model}) with the parameters in {\bf Set B},
initialised with the variational stationary state of the averaged
system. The first 4 frames show the managed evolution at the indicated times; the last frame shows the same initial data with the modulation switched off ($\gamma_1=0$) shortly before the numerical solution has to be terminated.}
\label{fig:4}
\end{figure}
\end{center}

Figure~\ref{fig:4} shows the density $|u|^2$ of the \emph{original} model of
Set~B, initiated with the variationally found stationary state of the averaged system. In the first 4 frames the management is on; the wave packet keeps a localised core over
many internal periods, breathing about the stationary width, and loses only
the fraction of its norm that reaches the absorbing layer. In the last frame
the same initial data are propagated with $\gamma_1=0$, i.e.\ with the mean
coupling $\gamma_0$ alone: since $N_0>N_c$ the packet collapses, and the run
has to be terminated. The two rows are plotted on the same colour scale up to
the last frame, where the unmanaged density has grown by more than an order of
magnitude. This is the direct numerical statement of the result of
Sec.~\ref{sec:noblowup}: the same initial data, the same mean nonlinearity,
and the modulation decides between collapse and a long-lived soliton.

\begin{center}
\begin{figure}[H]
\includegraphics[scale=0.6]{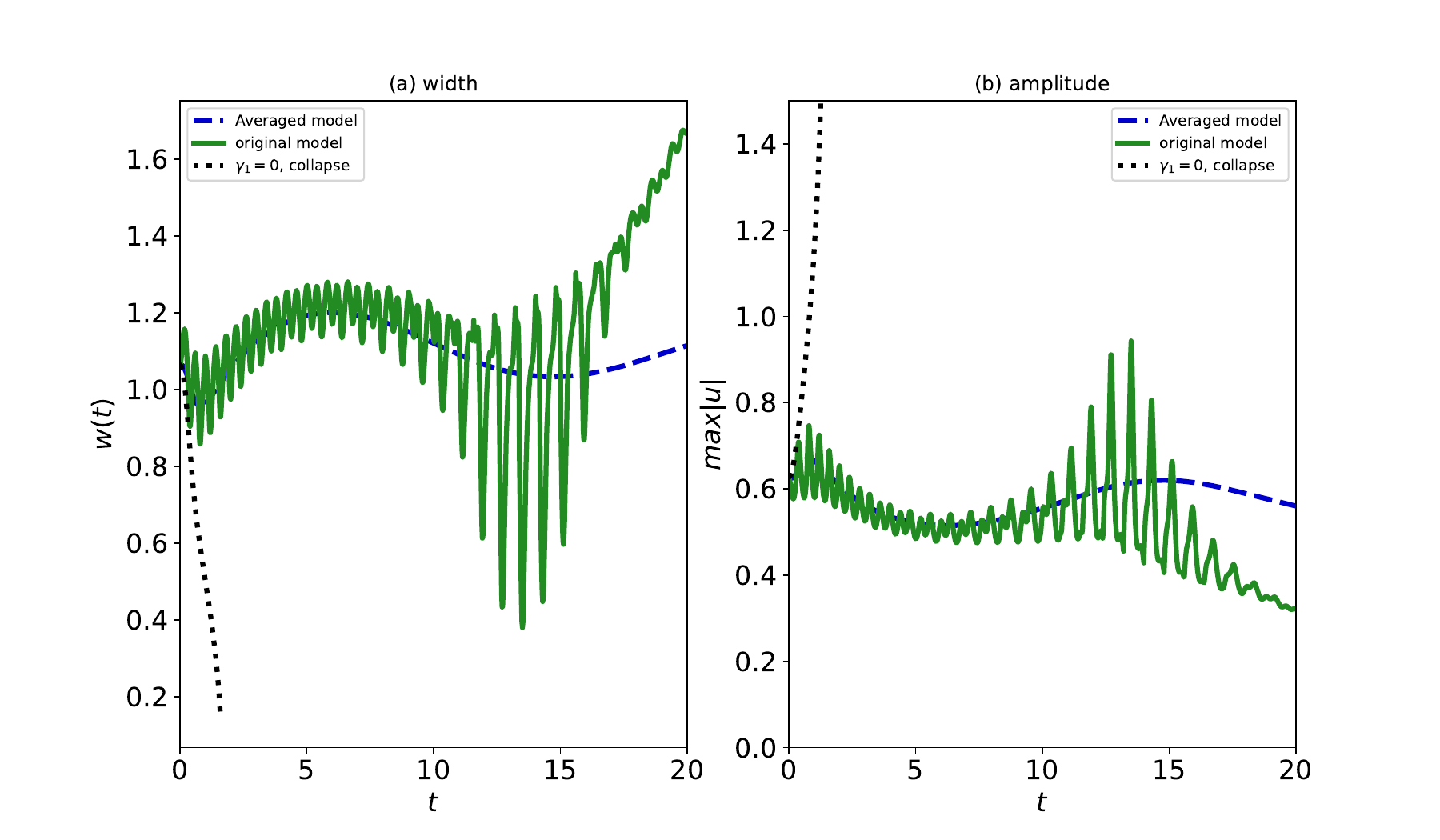}
\caption{(Color online) Time evolution of (a) the width $w(t)$ and (b) the
peak amplitude $\max|u|(t)$ for the same run as in Figure~\ref{fig:4}
(Set~B, initial data equal to the variational stationary state of the
averaged system). Green solid: original (managed) model; blue dashed:
prediction of the averaged model; black dotted: unmanaged evolution with
$\gamma_1=0$, which collapses shortly after $t\approx 1$.}
\label{fig:5}
\end{figure}
\end{center}

A complementary, quantitative view of the same run is given in
Figure~\ref{fig:5}, which tracks the width $w(t)$ and the peak amplitude
$\max|u|(t)$ of the wave packet instead of its full spatial profile. The
green curve is the original (managed) model: both observables oscillate
rapidly at the modulation frequency while their envelope follows the slowly varying prediction of the averaged model (blue dashed curve) of
Sec.~\ref{sec:noblowup}. Over the interval $t\in[0,15]$ shown, both
$w(t)$ and $\max|u|(t)$ stay bounded and of order unity, confirming that
the packet keeps breathing rather than collapsing. The black dotted curve
shows the same initial data evolved with $\gamma_1=0$: the width collapses
to zero and the amplitude diverges within about one modulation period, and
the run is stopped once the numerical resolution can no longer be trusted.
Taken together with Figure~\ref{fig:4}, this shows in terms of the scalar diagnostics used throughout the paper that it is specifically the
management, and not some accidental feature of a particular snapshot, that
arrests the collapse.

\subsection{Asymmetric solitons}\label{sec:asymnum}

Set~C is the asymmetric case: all three interactions attractive,
$N_v/N_u=5$, and --- what makes the vector nature visible --- two components
of clearly different size. The design follows Sec.~\ref{sec:asymmetric}: the
width ratio on the existence boundary is
$\kappa_*=(a_2/a_1)^{1/4}$, so the asymmetry is controlled by the contrast
between $g_{11}N_u$ and $g_{22}N_v$ and not by the population imbalance alone
(for $g_{11}=g_{22}$ one finds $\kappa_*\simeq1$ even at $N_v/N_u=1.5$). Here
$a_1=0.196$, $a_2=1.682$, $a_{12}=-3.2$, so that
$\rho\equiv(\sqrt{a_1}+\sqrt{a_2})^2/(-a_{12})=0.945<1$ --- the state exists, according to Eq.~(\ref{eq:asymbound}) --- and $\kappa_*=1.711$.

Newton iteration on $\nabla U=0$ converges from the seed $(0.7,1.2)$ in five
steps to $|\nabla U|<10^{-13}$ and gives
\begin{equation}\label{eq:Cfixed}
w_u=0.768,\quad w_v=1.29,\quad
A_u=0.656,\quad A_v=0.874,\quad \bar H=-0.05 ,
\end{equation}
with $w_v/w_u=1.1.677$ and $A_v/A_u=1.33$. The Hessian gives
$\Omega_-=0.0.284$ and $\Omega_+=2.3$, both real, so the state is a genuine
minimum of $U$.

\begin{center}
\begin{figure}[H]
\includegraphics[scale=0.35]{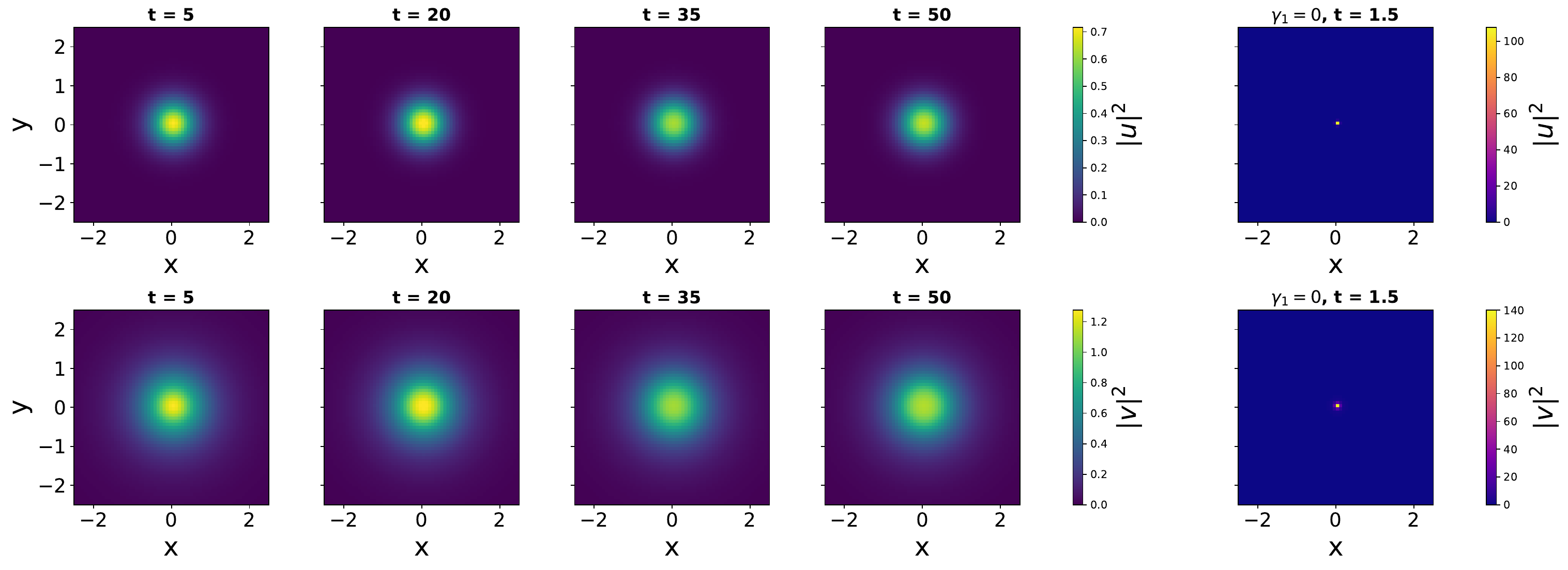}
\caption{(Color online) Density $|u|^2$ (top row) and $|v|^2$ (bottom row) of
the full model (\ref{eq:original_model}) with the parameters of {\bf Set C},
initialised with the fixed point (\ref{eq:Cfixed}). The first 4 columns show
the managed evolution at the indicated times; the last column shows the same
initial data with the modulation switched off ($\gamma_1=0$), shortly before
the run has to be terminated at $t=1.5$ due to collapse.}
\label{fig:6}
\end{figure}
\end{center}

As in Sec.~\ref{sec:arrest}, this variational prediction is checked against a
direct simulation of the full model. Figure~\ref{fig:6} shows the density
$|u|^2$ (top row) and $|v|^2$ (bottom row) of Set~C, initiated with the fixed
point (\ref{eq:Cfixed}). In the first 4 columns the management is on: both
components keep a single, well-localised peak over several internal periods. In the last column the same initial data are
propagated with $\gamma_1=0$: both components collapse together, on the
common time scale set by the three attractive couplings, and the run has to
be stopped at $t=1.5$.

\begin{center}
\begin{figure}[H]
\includegraphics[scale=0.6]{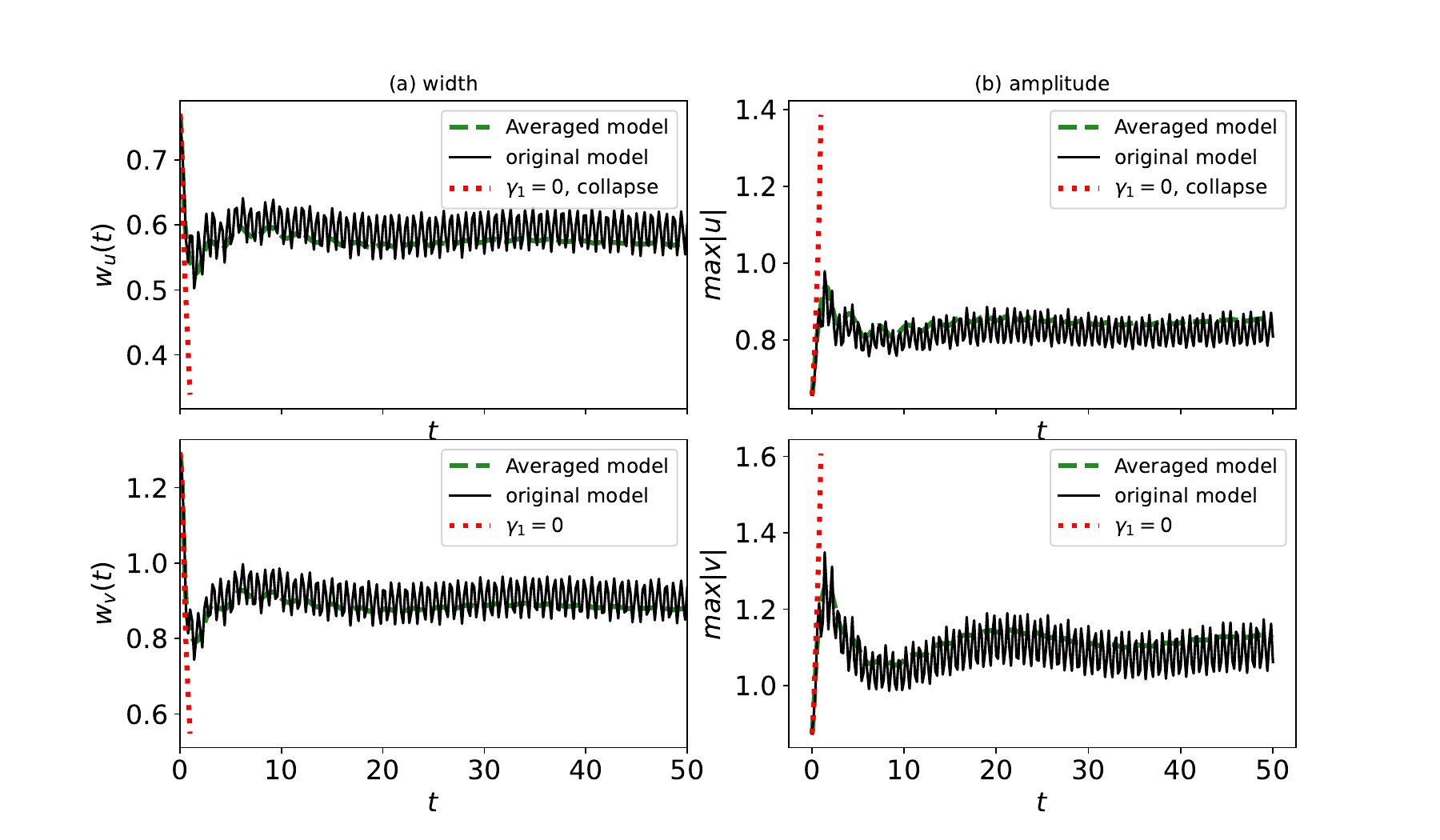}
\caption{(Color online) Time evolution of the widths $w_u(t)$, $w_v(t)$ (a)
and peak amplitudes $\max|u|(t)$, $\max|v|(t)$ (b) for the same run as in
Figure~\ref{fig:6} (Set~C, initial data equal to the fixed point
(\ref{eq:Cfixed})). Top row: $u$ component; bottom row: $v$ component. Black
solid: original (managed) model; green dashed: prediction of the averaged
model; red dotted: unmanaged evolution with $\gamma_1=0$, which collapses
shortly after $t\approx 1.5$.}
\label{fig:7}
\end{figure}
\end{center}

Figure~\ref{fig:7} gives the corresponding scalar diagnostics --- the widths
$w_u(t)$, $w_v(t)$ (panel a) and the peak amplitudes $\max|u|(t)$,
$\max|v|(t)$ (panel b) --- for the same run. For both components the
original model (green) oscillates about the averaged-model prediction (blue
dashed). The black dotted curve is again the $\gamma_1=0$ run: both
$w_u$ and $w_v$ fall away while $\max|u|$ and $\max|v|$ both diverge within
$t\lesssim1.5$ --- the same collapse already seen in the last column of
Figure~\ref{fig:6}. Management therefore prolongs the lifetime of the stationary state in the
asymmetric, two-component setting exactly as it did for the single-component, Set~B of Sec.~\ref{sec:arrest}.

\subsection{Estimates for an experiment}\label{sec:experiment}

It is useful to translate the invariant combinations ({\bf Set B}) into
physical units for a $^{39}$K spin mixture, in which the inter-species
scattering length is broadly tunable~\cite{dErrico,Semeghini}. For a transverse
trap frequency $\omega_z=2\pi\times100$~Hz one has
$l_z=\sqrt{\hbar/(m\omega_z)}=1.61\,\mu$m and a time unit
$\omega_z^{-1}=1.59$~ms. From the definition $g_{ij}=2\sqrt{2\pi}a_{ij}/l_z$ of
(\ref{eq:dim_units}), the existence condition
$N_0>N_c^{\rm ex}=\|R\|^2_{L^2}/(2|g+\gamma_0|)$ becomes
\begin{equation}\label{eq:expcond}
N_0\,|a_{11}+a_{12}|\ >\ \frac{\|R\|^2_{L^2}}{4\sqrt{2\pi}}\,l_z
\simeq1.167\,l_z=1.88\,\mu\mathrm{m},
\end{equation}
i.e. $N_c^{\rm ex}\simeq1.8\times10^{3}$ atoms per component for
$|a_{11}+a_{12}|\simeq20a_B$ ($a_B$ is the Bohr radius); with $\gamma_0=-2\pi g$
this corresponds to $a_{11}\simeq+3.8\,a_B$ and $a_{12}\simeq-23.8\,a_B$.
Our reference set $(g+\gamma_0)N_0=-2.35\pi$ then corresponds to
\begin{equation}
N_0\simeq2.2\times10^{3}\ \text{atoms per component}\quad
(N_0/N_c^{\rm ex}=1.26).
\end{equation}
Only the products $g_{ij}N_0$ are fixed by the theory, so the modulation
amplitude must be read off in the same way: $\gamma_1N_0$ and
$(g+\gamma_0)N_0$ share the same atom number, and their ratio gives
\begin{equation}
\Delta a_{12}=\frac{\gamma_1}{|g+\gamma_0|}\,|a_{11}+a_{12}|
\simeq5.0\ \mathrm{nm}\simeq95\,a_B ,
\end{equation}
and a modulation frequency $\omega\omega_z/2\pi\simeq1.6$~kHz. Modulation
amplitudes of order $10^2a_B$ are available near a Feshbach
resonance, and $\sim 1$~kHz modulation frequencies are accessible with optically induced
Feshbach resonances~\cite{Feshbach}.

\section{Conclusion}\label{Sec_Conclusion}
We have studied two-dimensional matter-wave vector solitons in a two-component BEC with attractive intra- and inter-species interactions, stabilized by strong nonlinearity management (NM) of the inter-species scattering length. Averaging the coupled Gross-Pitaevskii equations over the fast modulation produces an effective nonlinear quantum pressure term, proportional to the density of the other component, which for the {\it averaged} system rigorously excludes collapse: this term grows faster ($L^{-6}$) than the destabilizing mean-field terms ($L^{-2}$), bounding the averaged energy from below. It is important to stress, however, that this rigorous result holds for the averaged model only - a rigorous treatment of the original, explicitly time-dependent system remains an open problem, and the direct numerical simulations show that management does not eliminate collapse outright but substantially {\it prolongs the lifetime} of the localized state.

A variational analysis of the averaged system yields stationary widths, amplitudes, and normal-mode (breathing) frequencies for both symmetric and asymmetric solitons, in close agreement with numerically exact stationary states. Simulations of the full, unaveraged Gross-Pitaevskii equations, initialized at these stationary states, confirm this picture directly: with the modulation on, both components remain localized and simply breathe about the predicted width over many internal periods, whereas the identical initial data, propagated with the modulation switched off, collapse to a singularity within about one modulation period. Management therefore acts not as a permanent cure but as a mechanism that repeatedly interrupts the collapse dynamics, keeping the vector soliton alive far longer than the unmanaged condensate survives.

Translated into physical units for a $^{39}$K spin mixture, these solitons correspond to realistic experimental parameters — atom numbers of order $10^3$ per component, scattering-length modulation of order $10^2$ Bohr radii, and modulation frequencies around $1.6 \, kHz$ — all accessible with current optically induced Feshbach-resonance techniques, making the observation of long-lived, NM-prolonged two-dimensional vector solitons feasible with existing experimental capabilities.

\end{document}